# Low-Frequency Raman Spectroscopy of Few-Layer 2H-SnS$_2$

**Tharith Sriv[1,2], Kangwon Kim[1] and Hyeonsik Cheong[1,*]**

[1]Department of Physics, Sogang University, Seoul 04107, Korea
[2]Department of Physics, Royal University of Phnom Penh, Cambodia
*Corresponding author: hcheong@sogang.ac.kr

**ABSTRACT**

We investigated interlayer phonon modes of mechanically exfoliated few-layer 2H-SnS$_2$ samples by using room temperature low-frequency micro-Raman spectroscopy. Raman measurements were performed using laser wavelengths of $441.6, 514.4, 532$ and $632.8$ nm with power below $100\ \mu\text{W}$ and inside a vacuum chamber to avoid photo-oxidation. The intralayer $E_\text{g}$ and $A_\text{1g}$ modes are observed at $\sim 206$ cm$^{-1}$ and $314$ cm$^{-1}$, respectively, but the $E_\text{g}$ mode is much weaker for all excitation energies. The $A_\text{1g}$ mode exhibits strong resonant enhancement for the $532$ nm ($2.33$ eV) laser. In the low-frequency region, interlayer vibrational modes of shear and breathing modes are observed. These modes show characteristic dependence on the number of layers. The strengths of the interlayer interactions are estimated by fitting the interlayer mode frequencies using the linear chain model and are found to be $1.64 \times 10^{19}$ N $\cdot$ m$^{-3}$ and $5.03 \times 10^{19}$ N $\cdot$ m$^{-3}$ for the shear and breathing modes, respectively.

**Introduction**

Interest in two-dimensional (2D) materials such as hexagonal boron nitride (hBN), black phosphorus (BP) and transition-metal dichalcogenides (TMDs) since the discovery[1] of graphene in 2004 has significantly increased due to their unique structures and properties. Most TMD materials such as MoS(e)$_2$ and WS(e)$_2$ are indirect band gap semiconductors with band gap energies in the visible range but become direct in the monolayer limit[2-6]. Recently, tin disulfide (SnS$_2$) has attracted much interest because it is recognized as earth-abundant, relatively cheap and low-toxic material. Additionally, it has been shown to have high on/off current ratios for field effect transistors[7-8], fast photodetection[9] suitable for flexible photodetectors from UV to IR[10], interesting gas sensing property[11], and high optical absorption and photovoltaic activities[12].

SnS$_2$ is among the most important sulfide compounds of tin[13-14] and has more than 70 polytypes[13-15] differing from one another by stacking sequences of the individual sandwiched layers. The most common one is 2H-SnS$_2$ whose basic layer consists of a sheet of close-packed tin atoms sandwiched between two sheets of sulfur atoms[16-32]. It should be noted that in the literature, SnS$_2$ with a structure identical to that of 1T-MoS$_2$ as shown in Fig. 1(a) is called 2H-SnS$_2$[17-32], which should not be confused with the structure of 2H-MoS$_2$. In 2H-SnS$_2$, a metal atom is octahedrally coordinated by sulfur atoms, whereas the metal atom in 2H-MoS$_2$ possesses trigonal prismatic coordination[16]. Monolayers of 2H-SnS$_2$ are stacked exactly on top of one another to form 2H-polytype of bulk SnS$_2$. Between the neighboring layers of 2H-SnS$_2$, there exists weak van-der-Waals interaction[33] offering easy mechanical cleavage along the $c$-axis down to monolayer. Bulk 2H-SnS$_2$ belongs to the symmetry group of $D_{3d}^3$ ($P\bar{3}m1$) and has a trigonal structure with the lattice constants of $a = 3.6486$ Å and $c = 5.8992$ Å[17]. Unlike most TMDs, 2H-SnS$_2$ is projected to remain an indirect band gap semiconductor for all thicknesses, with the band gap ranging between 2.18 eV (bulk) and 2.41 eV (monolayer). Although chemical vapor deposition[34] and molecular beam epitaxy[35] growths have been tried, large-area growth of few-layer SnS$_2$ has not been realized yet. At the moment, mechanical exfoliation from bulk crystals yields the highest quality few-layer samples.

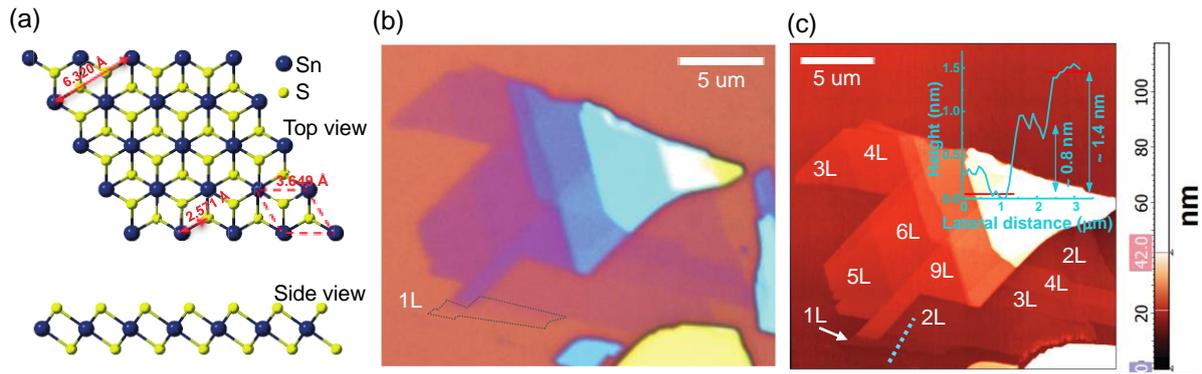

**Figure 1**. (a) Crystal structure of monolayer 2H-SnS$_2$. (b) Optical and (c) atomic force microscope (AFM) images of a mechanically exfoliated few-layer 2H-SnS$_2$ sample on a SiO$_2$/Si substrate.

Raman spectroscopy is one of the most widely used characterization tools for 2D layered materials to determine the number of layers as well as polytypes or strain effects. More importantly, one can use low-frequency Raman spectroscopy to study the interlayer interactions of few-layer materials by measuring the in-plane (shear) and out-of-plane (breathing) modes in the low-frequency region (<50 cm$^{-1}$). In the literature, the measured data of the shear and breathing modes are used to estimate the interlayer spring constants of the studied materials such as MoS$_2$ and WSe$_2$[36], MoSe$_2$[37], MoTe$_2$[38], WS$_2$[39], ReS(e)$_2$[40], Bi$_2$Te$_3$ and Bi$_2$Se$_3$[41], black phosphorus[42], and graphite[43-44] by fitting the experimental data to the linear chain model (LCM). Additionally, Luo et al. reported that the stacking sequence determines Raman intensities of observed interlayer shear modes[45]. However, experimental work on Raman properties of few-layer 2H-SnS$_2$ remains lacking although results for less-common 4H-SnS$_2$ have been reported[19]. The Raman spectrum of bulk 2H-SnS$_2$ shows two phonon modes at 315 cm$^{-1}$ ($A_{1g}$) and 205.5 cm$^{-1}$ ($E_g$), while that of 4H-SnS$_2$ shows several more modes[26]. This offers a clear distinction between 2H- and 4H-SnS$_2$. For few-layer SnS$_2$, Yuan et al.[19] recently reported a Raman study on mechanically-exfoliated monolayer and few-layer as well as bulk 4H-SnS$_2$. Nevertheless, low-frequency shear and breathing modes are not considered, i.e., interlayer interactions of the material remains uncovered in the Raman studies of few-layer $SnS_2$. In this work, we investigate the Raman spectra of mechanically-exfoliated few-layer $2H-SnS_2$ using four excitation energies. We also analyze the low-frequency Raman spectra to investigate the interlayer interaction in few-layer $2H-SnS_2$.

**Results and Discussion**

Figures 1(b) and (c) show the optical and AFM images of a $2H-SnS_2$ sample, respectively. The dotted outline in Figure 1(b) indicates where monolayer (1L) is located. The AFM measurements of this sample indicate the presence of several thicknesses as indicated. The 1L $2H-SnS_2$ has a thickness of ~0.6 nm[19]. Our AFM results show a step size of ~0.8 nm for 1L and ~1.4 nm for 2L, which is reasonable as there usually is a small extra thickness for the first layer in AFM measurements of 2D materials. This is either due to trapping of absorbed $H_2O$ molecules between the $2H-SnS_2$ and the $SiO_2/Si$ substrate[19] or imperfect adhesion of the sample on the substrate. We measured multiple sets of samples with thicknesses ranging from 1L to 14L and bulk. It is worth mentioning that no sign of degradation was observed after our few-layer $2H-SnS_2$ samples had been left in ambient condition for several weeks, but AFM measurements performed few hours after being exposed to the laser beam in the Raman measurements in ambient air showed degradation caused by photo-oxidation (see Supplementary Information). We therefore carried out all Raman measurements with the sample kept inside a vacuum chamber.

Figure 2(a) shows the low- and high-frequency Raman spectra of 5L $2H-SnS_2$ measured with four excitation energies. Vertical dashed-lines are guides for the eye. It is seen that the Raman signals are strongest for the 2.33 eV (532 nm) excitation laser. The out-of-plane $A_{1g}$ mode at ~314 $cm^{-1}$ is most prominent. The $E_g$ mode at ~206 $cm^{-1}$ is extremely weak and is barely resolved only in the spectrum taken with the excitation energy of 2.81 eV (441.6 nm). In the low-frequency region, the interlayer vibrational modes of in-plane shear (S) and out-of-plane breathing (B) modes are identified. Figure 2(b) shows the excitation energy dependence of the $A_{1g}$ mode for 1L to 14L 2H-SnS$_2$. The 532 nm (2.33 eV) excitation laser provides the strongest intensity of the $A_{1g}$ mode, which implies that the band gap of few-layer 2H-SnS$_2$ may be smaller than the recent theoretical prediction of 2.41 eV for 1L[16]. Figure 2(c) shows the dependence of the Raman spectrum on the number of layers. In addition to the $A_{1g}$ and $E_g$ modes, two other weak signals from $A_{1u}$ and $A_{1g}$-LA ($M$) modes are observed for bulk or thick samples at ~353 cm$^{-1}$ and ~140 cm$^{-1}$, respectively. The $A_{1u}$ mode is an infrared mode but appear probably due to activation by lattice disorders, whereas the two-phonon scattering[46-47] signal of $A_{1g}$-LA ($M$) is weak due to the small scattering cross section. Figure 2(d) shows the $E_g$ mode measured with the 441.6 nm excitation laser in cross polarization configuration since this excitation laser provided relatively stronger signals for the $E_g$ mode. No clear shift is observed as the thickness increases. Figure 2(e) indicates the evolution of the Raman intensity and the peak position of the $A_{1g}$ mode as a function of the number of layers. The error bars indicate the spectral resolution of the setup. The intensity of the $A_{1g}$ mode evolves monotonically with the number of layers up to ~11L. This mode also shows a slight blue-shift from 1L to 3L, which is in good agreement with recent theoretical results[16].

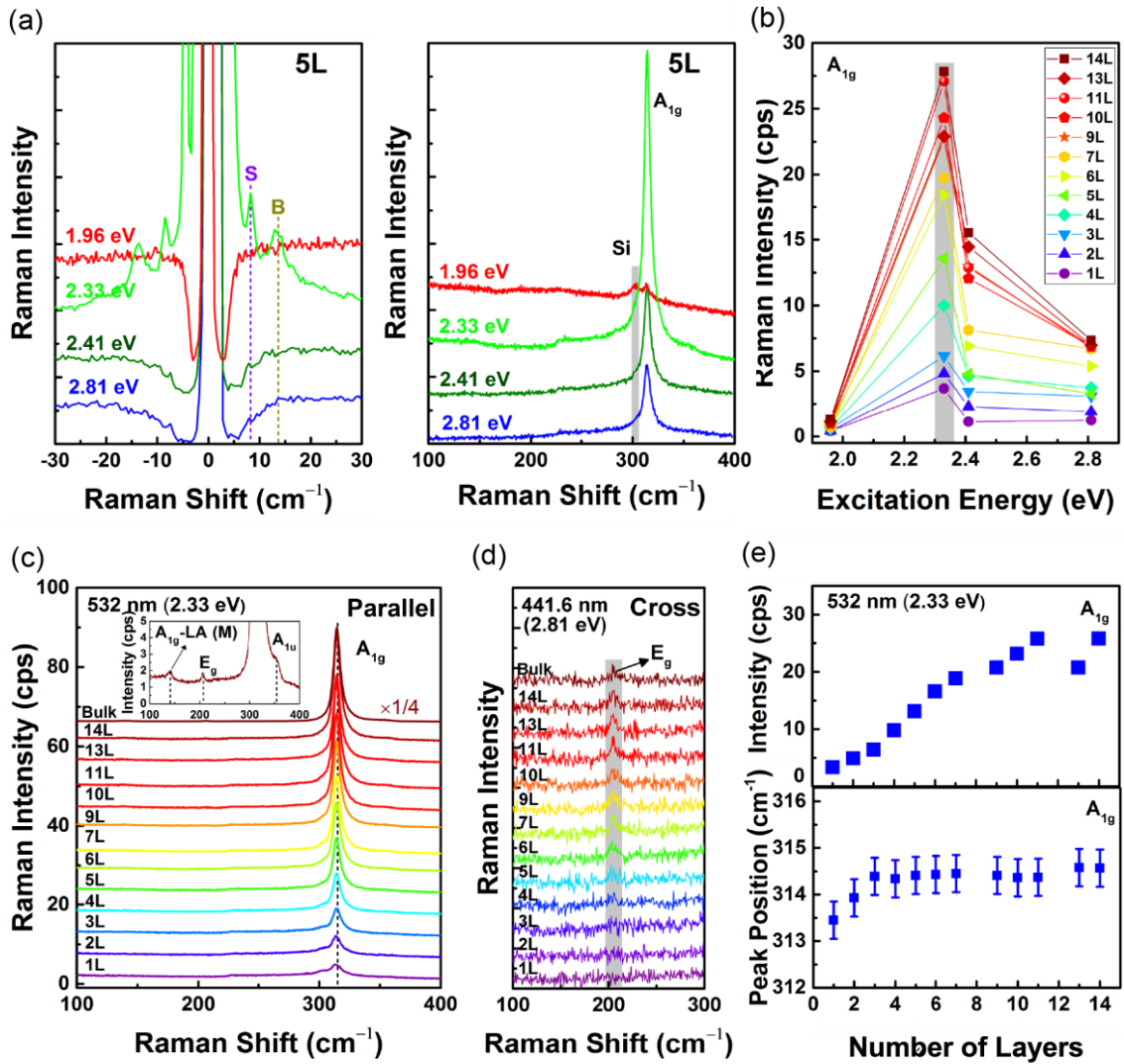

**Figure 2**. Raman spectra of few-layer 2H-SnS$_2$. (a) Low- and high-frequency modes of 5L 2H-SnS$_2$ measured with 441.6 , 514.4 , 532 and 632.8 nm lasers. (b) Excitation-energy dependence of the intensity of the $A_{1g}$ mode for 1L to 14L. (c) High-frequency modes of few-layer 2H-SnS$_2$ measured by using the 532 nm laser in parallel polarization configuration. The Raman intensity of $A_{1g}$ mode in bulk layer is multiplied by 1/4. Inset shows the $A_{1g}$-LA (M), $E_g$, and $A_{1u}$ modes of bulk 2H-SnS$_2$. (d) The $E_g$ mode measured by using the 441.6 nm laser in cross polarization configuration. (e) Evolution of the Raman intensity and peak positions of the $A_{1g}$ mode as a function of number of layers. The error bars indicate the spectral resolution of the setup.

For 1L 2H-SnS$_2$, there exist nine vibrational modes at the center of the Brillouin zone at the $\Gamma$ point: $\Gamma = A_{1g} + E_g + 2A_{2u} + 2E_u$[20,26]. Among six optical phonon modes, there are three Raman active modes ($A_{1g}$ and $E_g$) and three infrared-active modes ($A_{2u}$ and $E_u$). The three acoustic modes belong to $A_{2u}$ and $E_u$. The Raman scattering intensity is proportional to $|e_i \cdot \tilde{R} \cdot e_s|$, where $e_i$ represents the polarization vector of the incident light, $e_s$ that of the scattered light, and $\tilde{R}$ the Raman tensor. The Raman tensors can be expressed as[48],

$$A_{1g} = \begin{pmatrix} a & 0 & 0 \\ 0 & a & 0 \\ 0 & 0 & b \end{pmatrix} \text{ and } E_g = \begin{pmatrix} c & 0 & 0 \\ 0 & -c & d \\ 0 & d & 0 \end{pmatrix}, \begin{pmatrix} 0 & -c & -d \\ -c & 0 & 0 \\ -d & 0 & 0 \end{pmatrix} \tag{1}$$

In the backscattering geometry with the laser propagating in the $z$ direction, only the $E_g$ mode is observable in cross polarization, whereas both the $A_{1g}$ and $E_g$ modes can be observed in parallel polarization configuration. For the low-frequency interlayer modes that exist in 2L or thicker 2H-SnS$_2$, the shear modes correspond to $E_g$ and the breathing modes $A_{1g}$. By using polarized Raman measurements, one can thus distinguish shear and breathing modes unequivocally.

Figure 3(a) illustrates the polarization dependence of the Raman spectrum of 5L 2H-SnS$_2$. As a function of the relative scattering polarization angle with respect to the incident polarization direction, the intensities of the intralayer $A_{1g}$ mode and the interlayer breathing modes are modulated, whereas the intralayer $E_g$ mode and the interlayer shear modes are independent of the scattering polarization, which is consistent with the Raman tensor analysis above. Figure 3(b) illustrates the vibrations of in-plane shear and out-of-plane breathing. The evolution of the low-frequency interlayer vibrational modes as a function of the number of layers is shown in Figure 3(c)-(e). The shear and breathing modes can be distinguished by using polarized Raman measurements as explained before. Figure 3(c) shows the shear modes measured in cross polarization, in which the breathing modes are suppressed. Up to 2 shear modes (S1 and S2) are identified, and their positions depend sensitively on the number of layers. Figure 3(d) shows similar spectra measured in parallel polarization. Here, both the shear and breathing modes are observed. By comparing with Figure 3(c), one can unambiguously identify the breathing modes (B1 and B2). Figure 2(e) summarizes the evolution of the interlayer vibrational modes as a function of the number of layers. Since the high-frequency intralayer modes show little dependence on the number of layers beyond 3L, low-frequency Raman analysis would be the most reliable method to determine the number of layers of few-layer 2H-SnS$_2$.

As the low-frequency interlayer modes reflect the strength of the interlayer interaction, one can estimate the interlayer spring constants in the in-plane and out-of-plane directions by analyzing the frequencies of the shear and breathing modes, respectively. In the linear chain model[36, 41,49], assuming that only interactions between nearest-neighbor layers are important and by neglecting the substrate and surface effects, the angular frequency of the $\alpha$-th shear (breathing) mode in $N$-layer 2H-SnS$_2$ is given by,

$$\omega_\alpha = \frac{1}{\pi c} \sqrt{\frac{K}{2\mu} \left[ 1 - \cos\left( \frac{(\alpha - 1)\pi}{N} \right) \right]}, \qquad (2)$$

where $\alpha = 2, 3, \dots, N$ ($\alpha = 1$ corresponds to the zero-frequency acoustic mode at $\Gamma$ point in the Brillouin zone), $c$ is the speed of light in vacuum, $K$ is the in-plane (out-of-plane) force constant, and $\mu = 2.63352 \times 10^{-26}$ kg·Å$^{-2}$ is the mass per unit area of monolayer of 2H-SnS$_2$. The in-plane ($K=K_S$) and out-of-plane ($K=K_B$) force constants per unit area can then be obtained by fitting the experimentally obtained peak frequencies of the shear and breathing modes, respectively, to equation (2). Table 1 compares the force constants per unit area of 2H-SnS$_2$ thus obtained with those of other layered materials found in the literature. The interlayer interaction in 2H-SnS$_2$ is significantly weaker than in most materials compared.

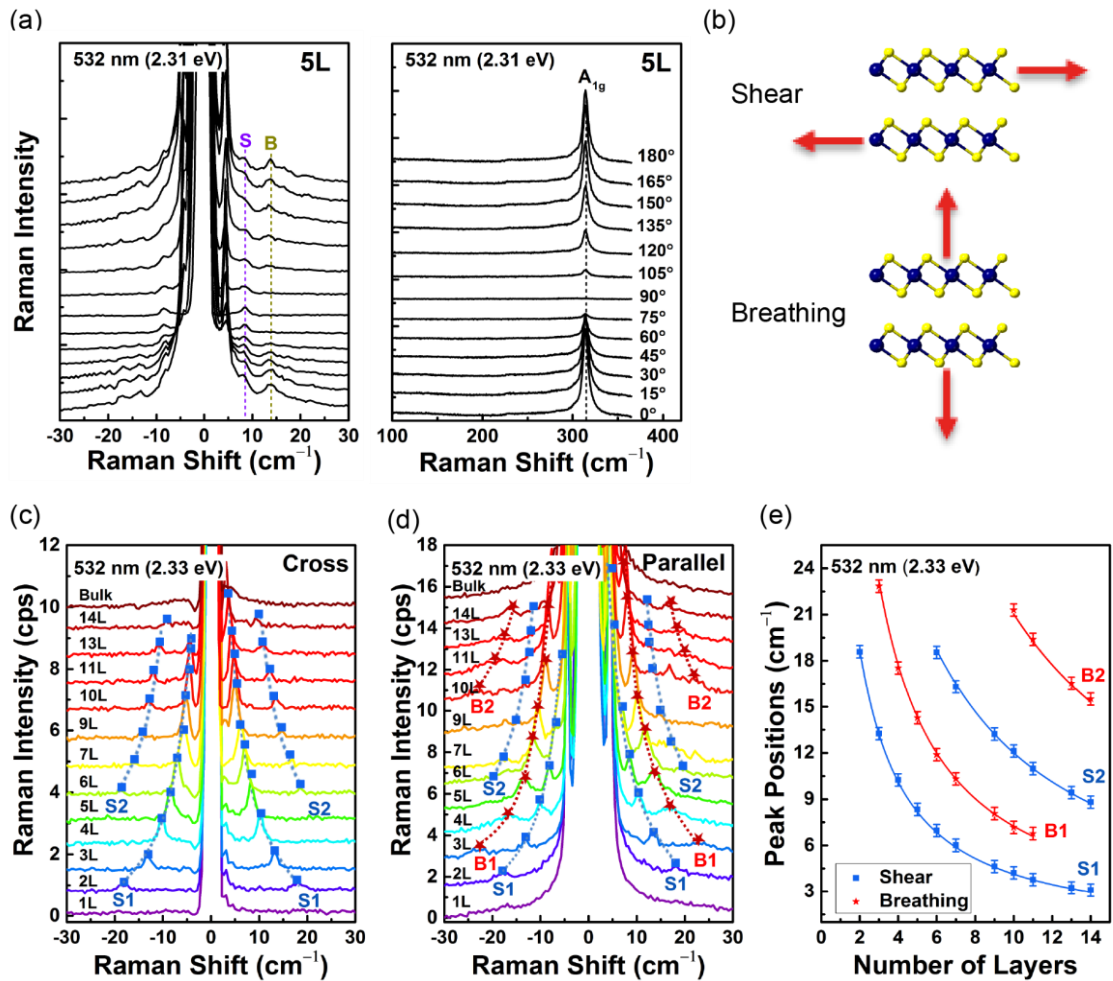

**Figure 3**. (a) Scattering angle dependence of the Raman spectra of 5L 2H-SnS$_2$ measured by using 2.33 eV (532 nm) excitation laser. Labels S (shear) and B (breathing) indicate the positions of shear and breathing modes resolved in 5L 2H-SnS$_2$, respectively. (b) Schematics of interlayer in-plane shear and out-of-plane breathing modes. (c) Shear modes measured in cross polarization. (d) Shear and breathing modes measured in parallel polarization. The dashed curves are guides for the eye. (e) Peak positions as a function of number of layers. Solid curves are fitting results using the linear chain model.

| Material | $K_S (10^{19}\ \text{N} \cdot \text{m}^{-3})$ | $K_B (10^{19}\ \text{N} \cdot \text{m}^{-3})$ |
|---|---|---|
| SnS$_2$ (this work) | 1.64 | 5.03 |
| MoS$_2$[36] | 2.72 | 8.62 |
| MoSe$_2$[37] | 2.92 | 8.73 |
| MoTe$_2$[38] | 3.44 | 7.83 |
| WS$_2$[39] | 2.99 | 9.10 |
| WSe$_2$[36] | 3.07 | 8.63 |
| ReS$_2$[40] | 1.71/1.89 | 6.90 |
| ReSe$_2$[40] | 1.78/1.94 | 6.90 |
| Bi$_2$Te$_3$[41] | 4.57 | 13.33 |
| Bi$_2$Se$_3$[41] | 2.27 | 5.26 |
| Black phosphorus[42] | - | 12.3 |
| Graphite[43-44] | 1.20 | 9.40 |

**Table 1**. Force constants per unit area of 2H-SnS$_2$ obtained by fitting experimental data to the linear chain model and comparison with those of other TMD materials.

In summary, we investigated lattice dynamics of mechanically-exfoliated few-layer 2H-SnS$_2$ by room temperature low-frequency micro-Raman spectroscopy using four different excitation energies. In monolayer, the intralayer out-of-plane $A_{1g}$ ($\sim$314 cm$^{-1}$) mode is most prominent, whereas in thick samples and bulk, the weak in-plane $E_g$ ($\sim$206 cm$^{-1}$) mode as well as two additional modes such as $A_{1g} - \text{LA}(M)$ ($\sim$140 cm$^{-1}$) and $A_{1u}$ ($\sim$353 cm$^{-1}$) are resolved. The 2.33 eV (532 nm) excitation laser provides the strongest Raman signals of intralayer $A_{1g}$ mode and interlayer shear and breathing modes, whereas the $E_g$ mode appears stronger for the 2.81 eV (441.6 nm) excitation. For the $A_{1g}$ mode, the Raman shift is slightly sensitive to thickness for 1L-3L, but not for thicker material. The shear and breathing modes show strong dependence on the thickness, which provides a robust criterion for determination of the thickness using Raman spectroscopy. The interlayer interactions obtained by analyzing the interlayer vibrational modes are weaker than in most other layered materials. These results provide valuable information on materials parameters for device designs using few-layer 2H-SnS$_2$.

## Methods

Few-layer 2H-SnS$_2$ samples were prepared from a SnS$_2$ single-crystal (HQ Graphene) onto SiO$_2$/Si substrates with 280 nm-thick oxide layer by mechanical exfoliation. The thickness of the samples was determined by atomic force microscope (AFM) and further confirmed by Raman measurements. The AFM measurements were performed by using a commercial AFM system (NT-MDT NTEGRA Spectra). Room temperature micro-Raman spectroscopy was conducted in backscattering geometry using four different excitation energies: the 441.6 nm (2.81 eV) line of a He-Cd laser, the 514.4 nm (2.41 eV) line of a diode-pumped laser (Cobolt), the 532 nm (2.33 eV) line of a diode-pumped solid-state (DPSS) laser, and the 632.8 nm (1.96 eV) line of a He-Ne laser. The input laser beam was focused onto the samples by a 40 × microscope objective lens (0.6 NA), and the scattered light was collected and collimated by the same objective lens. The laser of power below 100 μW was used. All measurements were performed with the sample in a vacuum chamber to prevent photo-oxidation. AFM images [Supplementary Information Figure S1] taken after Raman measurements confirmed that there were no apparent damages. Volume holographic filters (Ondax and OptiGrate) were used to access the low-frequency range below 50 cm$^{-1}$. The Raman scattering signals were dispersed by a Jobin-Yvon iHR550 spectrometer with a 2400 grooves/mm grating (400 nm blaze) and detected by a liquid-nitrogen-cooled back-illuminated charged-couple-device (CCD) detector. The spectral resolution was below 1 cm$^{-1}$.

## Acknowledgements

This work was supported by the National Research Foundation (NRF) grant funded by the Korean government (MSIT) (NRF-2016R1A2B3008363 and No. 2017R1A5A1014862, SRC Progream: vdWMRC center) and by a grant (No. 2011-0031630) from the Center for Advanced Soft Electronics under the Global Frontier Research Program of MSIT. T.S. acknowledges supports from the Korean Government Scholarship Program (KGSP), the International Science Programme (ISP), Uppsala University, Sweden and Royal University of Phnom Penh, Cambodia.

## Author contributions

H.C. conceived the experiments. T.S. and K.K. carried out measurements. All authors analyzed the data and wrote the manuscript.

## Additional information

**Supplementary Information** accompanies the paper at http://www.nature.com/srep.

The authors declare no competing interests.